\documentclass[%
 reprint,
superscriptaddress,
 amsmath,amssymb,
 aps,
prl
]{revtex4-2}

\usepackage{changes}
\usepackage{graphicx}
\usepackage{dcolumn}
\usepackage{bm}
\usepackage[colorlinks=true,linkcolor=blue,citecolor=blue,urlcolor=blue]{hyperref}

\begin{document}

\preprint{APS/123-QED}

\title{Controlling crystal cleavage in Focused Ion Beam shaped specimens for surface spectroscopy}

\author{A. Hunter}
\thanks{These authors contributed equally to this work.}
\affiliation{Department of Quantum Matter Physics, University of Geneva, 24 Quai Ernest-Ansermet, 1211 Geneva 4, Switzerland}

\author{C. Putzke}
\thanks{These authors contributed equally to this work.}
\affiliation{Max Planck Institute for the Structure and Dynamics of Matter, Hamburg, Germany}
\affiliation{Laboratory of Quantum Materials (QMAT), Institute of Materials (IMX), École Polytechnique Fédérale de Lausanne (EPFL), Lausanne, Switzerland}

\author{I. Gaponenko}
\affiliation{Department of Quantum Matter Physics, University of Geneva, 24 Quai Ernest-Ansermet, 1211 Geneva 4, Switzerland}

\author{A. Tamai}
\affiliation{Department of Quantum Matter Physics, University of Geneva, 24 Quai Ernest-Ansermet, 1211 Geneva 4, Switzerland}

\author{F. Baumberger}
\affiliation{Department of Quantum Matter Physics, University of Geneva, 24 Quai Ernest-Ansermet, 1211 Geneva 4, Switzerland}
\affiliation{Swiss Light Source, Paul Scherrer Institut, CH-5232 Villigen PSI, Switzerland}

\author{P.J.W. Moll}
\affiliation{Max Planck Institute for the Structure and Dynamics of Matter, Hamburg, Germany}
\affiliation{Laboratory of Quantum Materials (QMAT), Institute of Materials (IMX), École Polytechnique Fédérale de Lausanne (EPFL), Lausanne, Switzerland}

\newcommand{\SRO}{Sr$_2$RuO$_4$}
\newcommand{\STO}{SrTiO$_3$}

\date{\today}

\begin{abstract}
Our understanding of quantum materials is commonly based on precise determinations of their electronic spectrum by spectroscopic means, most notably angle-resolved photoemission spectroscopy (ARPES) and scanning tunneling microscopy (STM). Both require atomically clean and flat crystal surfaces which traditionally are prepared by \textit{in-situ} mechanical cleaving in ultrahigh vacuum chambers. We present a new approach that addresses three main issues of the current state-of-the-art methods: 1) Cleaving is a highly stochastic and thus inefficient process; 2) Fracture processes are governed by the bonds in a bulk crystal, and many materials and surfaces simply do not cleave; 3) The location of the cleave is random, preventing data collection at specified regions of interest. Our new workflow is based on Focused Ion Beam (FIB) machining of micro-stress lenses in which shape (rather than crystalline) anisotropy dictates the plane of cleavage, which can be placed at a specific target layer. As proof-of-principle we show ARPES results from micro-cleaves of \SRO{} along the \textit{ac} plane and from two surface orientations of \STO, a notoriously difficult to cleave cubic perovskite.
\end{abstract}

\maketitle

\section{Introduction}
The physics of quantum materials is dominated by the complex behavior of the interacting many-electron system encoded in the single particle spectral function. Angle resolved photoemission spectroscopy (ARPES) and scanning tunneling microscopy (STM) probe the spectral function in reciprocal and real space, respectively, and have played a major role in shaping our understanding of quantum materials~\cite{Sobota2021,Fischer2007}. Both techniques are highly surface sensitive and thus rely on the availability of clean and atomically flat surfaces.
Methods to prepare high quality surfaces have seen little development over the last two decades. Surfaces of elemental metals and some semiconductors can reliably be prepared by repeated sputter-anneal cycles. However, the same method cannot generally be used for binary or ternary quantum materials because the sputter yield is element specific and sputtering thus changes the stoichiometry in an uncontrolled way. Similarly, annealing can lead to preferential desorption of certain elements and may in addition promote undesired surface reconstructions.
Complex materials are thus typically cleaved in ultrahigh vacuum immediately prior to ARPES or STM measurements. This is most commonly done by gluing in air a small stick ('top post') on a piece of crystal with often irregular shape. After insertion in UHV via a load-lock system, a bending force is applied to the top-post, usually by touching it with a wobble-stick. In the ideal case, this cleaves the crystal and results in a sufficiently large and flat surface to be probed in experiment. While this method has been highly successful, it does have multiple drawbacks. 
i) Cleaving is a stochastic process strongly affected by the often irregular shape of crystals and by defects such as the intergrowth of different phases. This leads to a large variability in the data quality and may even cause experiments to fail.
ii) Many interesting materials such as intermetallics, cubic perovskites but also the layered La- and Tl-cuprates have nearly isotropic bond strengths and do not cleave well.
iii) The crystal breaks at the mechanically weakest link, which usually corresponds to an enhanced defect density.
iv) The location of a cleave cannot be controlled. This proved to be a severe limitation for ARPES experiments under uniaxial strain~\cite{Ricco2018,Sunko2019}. If straining forces are applied directly to a sample, the resulting strain field is inversely proportional to the cross-section and will thus vary over the sample surface unless the crystal shape remains highly regular after the cleavage process. A similar problem is observed in approaches where thin samples are not directly deformed, but rather glued to a substrate which is itself strained \textit{in-situ}.  With such a setup, the strain field  partially relaxes through the epoxy and the degree of relaxation further depends on the thickness of the crystal remaining after cleavage.

Alternative cleavage methods use mechanically actuated knife edges in fixed cleaving stations~\cite{Angot1991,Schmid2006,Muro2010,Sander2022} or on sample plates~\cite{Mansson2007}. Applying high clamping forces through macroscopic anvils proved powerful for cleaving the (001) surface of \STO, which has been attributed to strain-induced ferroelectricity~\cite{Sokolovic2019,Sokolovic2021}. Although promising for certain cases, these methods have restrictions with regards to the sample material, surface plane and do not allow micrometer precision in placing the cleavage plane.

\begin{figure*}[!ht]
\centering
\includegraphics[width=1\textwidth]{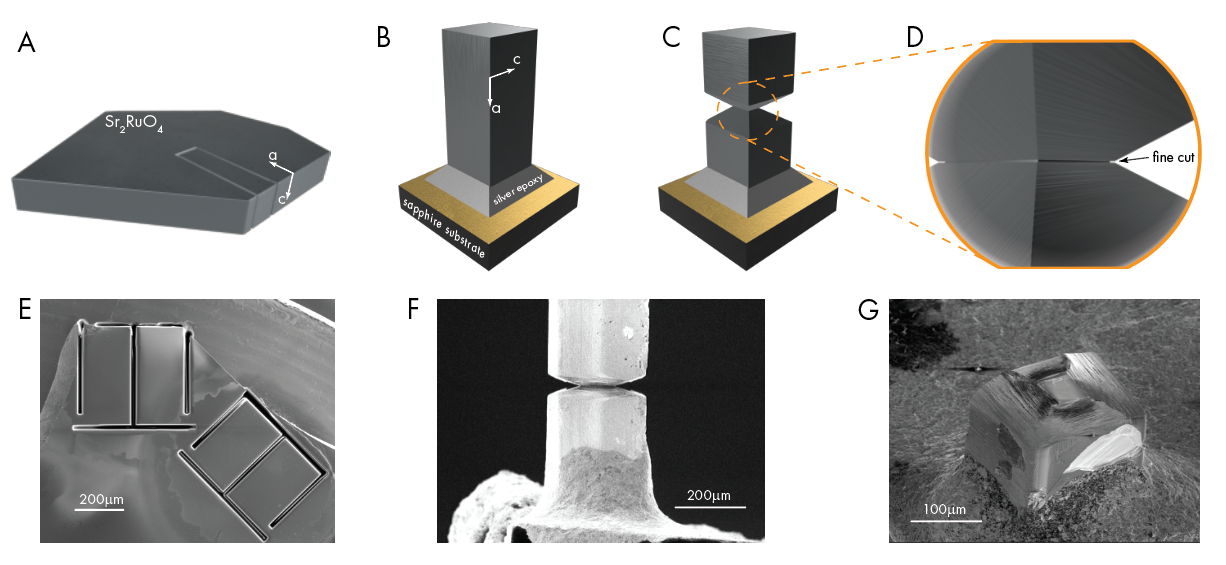}
\caption{(a) Pillars are cut from an aligned single crystal with a bridge maintaining a connection between the pillar and crystal. 
(b) The pillar is mounted upright on a gold coated sapphire substrate using silver epoxy (Epotech H20E) and a ceramic post or thin metal wire is glued on the top face to permit cleaving with standard tools available at ARPES beamlines. (c) The pillar is necked to a cross section of less than $80\times80$~$\mu$m$^2$. 
(d) The center of the neck is fine structured with a single line cut to produce a sub-micrometer strain lens in the desired cleaving plane. 
The second row depicts SEM micrographs corresponding to the sketches in (a-c) demonstrating the process of a \SRO{} pillar fabrication. 
(e) A \SRO{} crystal with cuts defining 2 pillars each with \textit{ac} and \textit{aac} cleavage plane corresponding to [100] and [110] surface normals, respectively. 
(f) SEM micrograph of the pillar mounted and necked.
(g) SEM micrograph after the top of the pillar was pushed off and the sample cleaved perpendicular to the atomic layers.}
\label{fig_FIB_workflow}
\end{figure*}

Here, we introduce a new approach based on Focused Ion beam (FIB) machining of strain lenses. We show that this permits the precise positioning of cleavage planes, allows cleaving of crystals along non-standard planes and results in high-quality surfaces suitable for ARPES experiments even in hard to cleave materials.  

\section{Sample preparation}

In order to achieve a reliable cleave in the desired direction we design the shape of the crystal such that the stress present during the cleave is focused on a specific crystallographic plane that is the target of investigation. To achieve this we use FIB mircostructuring. The specific workflow shown in Fig.~\ref{fig_FIB_workflow} depicts the microfabrication steps from a single crystal to a micropillar ready for cleaving. A Xenon-plasma FIB was chosen for this task to allow for rapid machining even on the macro-scale because of the high ion currents and fast milling rates this technology offers. The starting point is a single crystal, whose crystallographic axes were determined using Laue diffraction. Pillar shaped pieces are cut from the sample using an acceleration voltage of 30~kV and beam current of 2.5~$\mu$A. The goal of this step is to cut through the entire crystal depth. Such deep cuts are accelerated when the crystal is not mounted to a substrate but freely suspended in space.
 In this configuration, once the beam pierces through the bottom of the crystal, the sputtered material can leave the trench via the bottom. This greatly improves speed of material removal and thereby allows for higher aspect ratios of width to depth of the FIB cut. Practically, this free suspension is realized by mounting the crystal hanging over the edge of a silicon wafer, which also facilitates manual handling. Without direct substrate support, the pillar is hanging in free space and hence a thin bridge connecting it to the parent crystal is required to keep it in place during the fabrication process. 
After coarse cutting the pillar edges, one of the two small cross section surfaces is chosen as the future bottom surface, which is used to mount the pillar to the substrate in the next step. That surface is smoothened using a grazing incidence angle of the Xenon beam at a 200\,nA current. This ensures a perpendicular and mechanically rigid mount of the pillar on the substrate. It is important to keep the bridge intact during these steps to avoid the pillar to fall out of the parent crystal. At last, this bridge is then thinned to approximately $1\times1$~$\mu$m$^2$, which allows ex-situ mechanical removal of the pillar from the remaining crystal without damage. The pillar can then be poked out under an optical microscope by applying a weak force to it, breaking the bridge. This can be done either manually using a sharpened wooden stick or semi-automatically using a micromanipulator.

The free pillar is then mounted onto a conductive substrate by carefully applying silver epoxy (Epotech H20E) under an optical microscope. This ensures good electrical contact for further FIB machining as well as the ARPES measurements, which is important to avoid charging effects. Gold coated sapphire or silicon chips perform well for this task, yet any hard, conductive substrate is suitable.
Before further processing, we glue either a thin wire or a ceramic post again by silver epoxy on the top face of the pillar to facilitate cleaving with a wobblestick at ARPES beamlines.

Subsequently, the mounted pillar is transferred to a holder pre-tilted at $36^{\circ}$  to process the side planes in the same Plasma FIB system. This holder is used to enable FIB access to the sides of the now vertically mounted pillar.
A V-shaped cut is produced on both sides of the pillar by irradiating triangular milling patterns on both sides in rapid alteration, sometimes called a parallel milling mode. This ensures that the tips of the V-shaped patterns remain well aligned to each other even in case of drifts during the milling procedure. The coarse thinning is done with a Xenon ion beam current set to 2.5\,$\mu$A to reduce the cross section by 50 percent or about 100\,$\mu m$, and then sequentially reduced to 200\,nA and 60\,nA for another 10\,$\mu m$ on each side respectively. The pillar is then rotated by 90 degrees around its axis on the pre-tilted holder and the procedure is repeated. The V-shaped neck reduces the cross sectional area of the pillar to $80\times80\,\mu$m$^2$ or less. This neck thickness has offered a good tradeoff between the degree of strain lensing and the size of the cleaved surface. Naturally, this is material and goal specific yet the dimensions can be trivially readjusted. 

To finally define the cleavage plane precisely, a line cut is performed at a beam current of 0.7\,nA in the center of the neck. The line-cut pattern is extended beyond the sample to ensure that the line is cut in the surface facing the FIB column as well as the sides of the neck. The key aspect of the line pattern is that the ion beam is guided in one dimension only, thus directly tracing out the final cleaving plane. This cut is performed for about 1~min (See Fig.~\ref{fig_FIB_workflow} d). It provides the final placement of the strain lens and hence is only performed once on one face of the sample, as it is technically challenging to cut a second feature with high alignment precision to a previous one after a substantial rotation of such a large object. Cutting this line once ensures a uniquely defined cleavage plane. 

At this stage the pillar is ready for cleaving and any contact to the top of the pillar could lead to irreversible damage and accidental cleavage. Hence it is desirable to directly mount the substrate on the final holder for the ARPES chamber at this stage to avoid unnecessary sample manipulation later on. 

\section{Results and discussion}
\begin{figure*}[]
\centering
\includegraphics[width=0.9\textwidth]{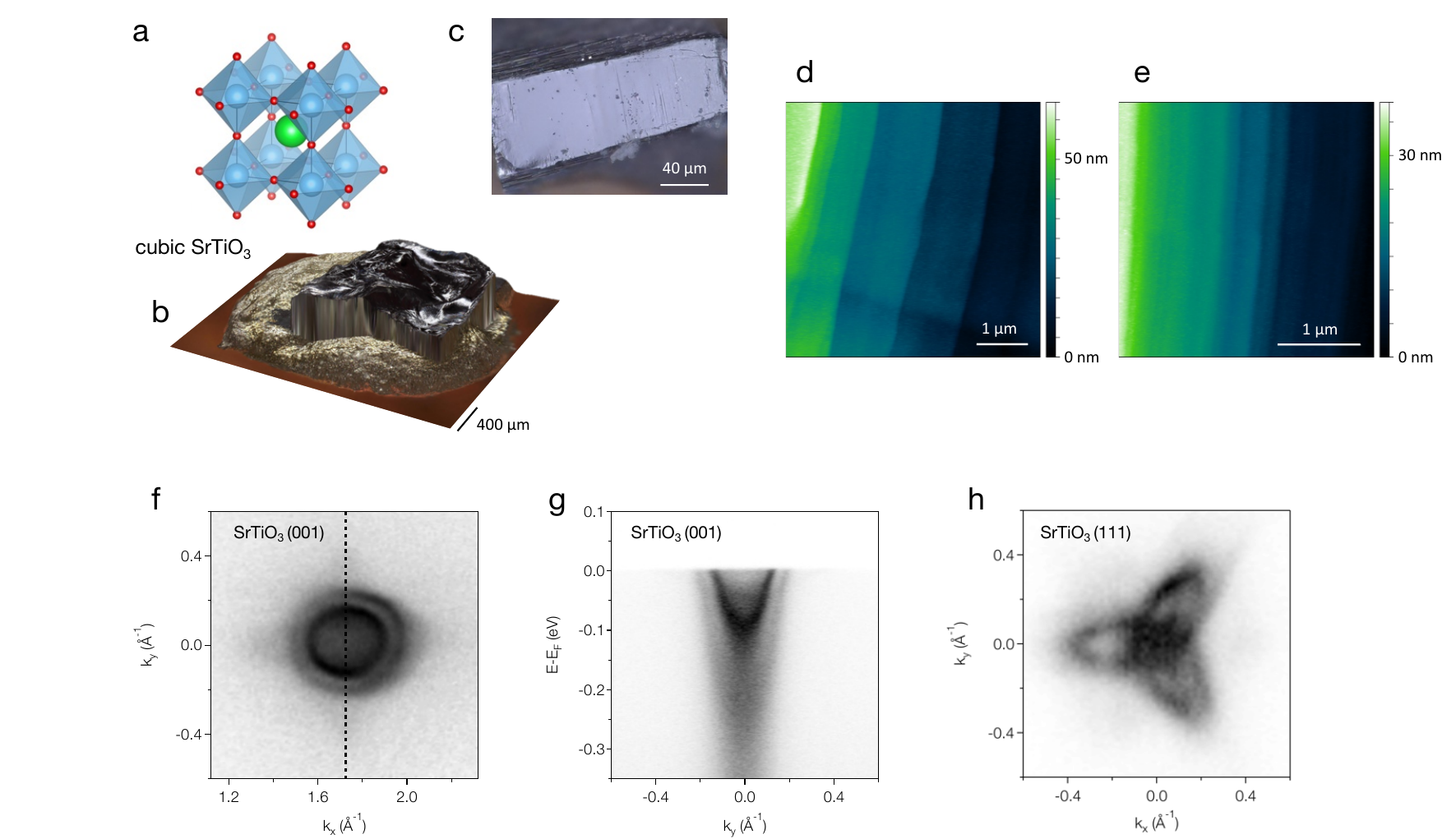}
\caption{First results from \STO{} micro-cleaves. (a) Structural model of \STO. (b) 3D optical profilometry of a conventionally cleaved/broken crystal. (c) optical micrograph of a cleaved \STO(001) micro-pillar. (d,e) AFM images of two different areas of the \STO(001) micro-pillar used for the ARPES measurements shown in (f,g) taken \textit{ex-situ} after the ARPES measurements. 
(f,g) Fermi surface and energy-momentum dispersion data of \STO(001) taken in the second Brillouin zone with $h\nu=47$~eV. (h) Fermi surface taken with $h\nu=108$~eV on the cleaved surface of a \STO(111) micro-pillar.}
\label{fig_STO}
\end{figure*}

\begin{figure*}[]
\centering
\includegraphics[width=\textwidth]{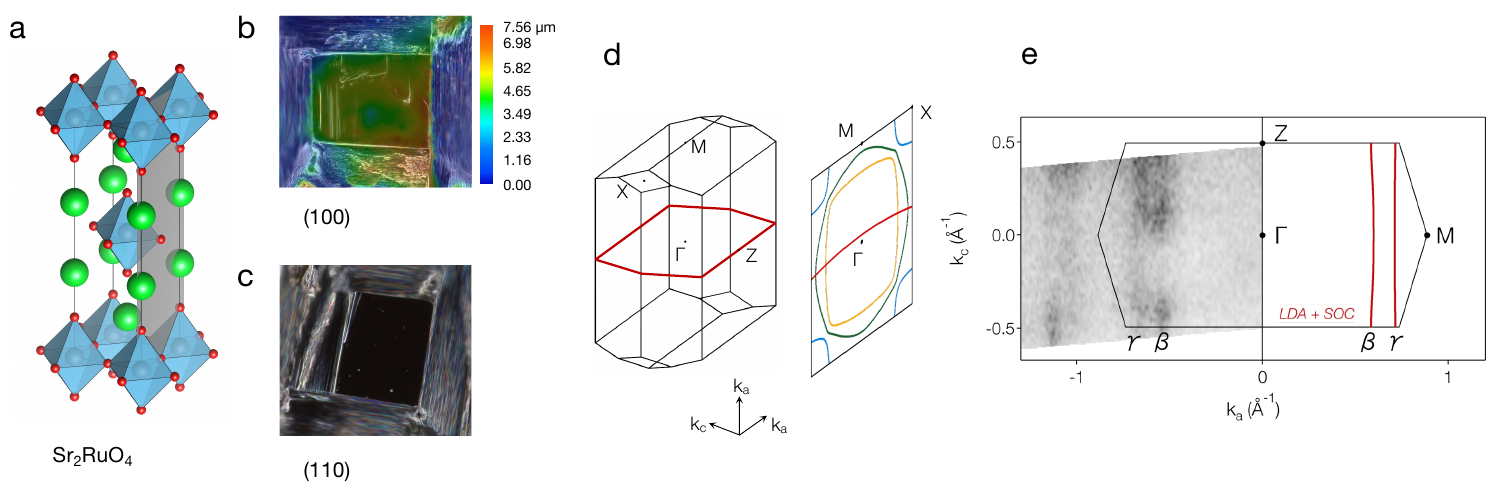}
\caption{c-axis cleaves of \SRO.
(a) Structural model with the \textit{ac} plane indicated in grey. (b) Optical profilometry of a cleaved (100) surface (\textit{ac} plane). (c) Optical micrograph of a \SRO(110) surface (\textit{aac} plane).
(d) Bulk Brillouin zone of \SRO{} with a schematic of the Fermi surface in the $(k_a,k_a)$ plane. 
The vertical momentum axis $k_a$ is perpendicular to the $ac$ surface. A photon energy of 83~eV probes the high-symmetry plane indicated in red.
(e) Fermi surface map from an \SRO(100) surface (\textit{ac} plane) taken with $h\nu=83$~eV. Red lines show the density functional theory Fermi surface calculated within the local density approximation (LDA) including spin-orbit coupling (SOC). }
\label{fig_SRO}
\end{figure*}

In Figs.~\ref{fig_STO},\ref{fig_SRO} we show first results from micro-cleaves of \STO{} and \SRO, two oxides with unusual electronic properties.
\STO{} attracts interest because of its catalytic and photo-catalytic activity and is widely used as substrate in oxide heteroepitaxy. 
The electronic structure of \STO{} gained renewed interest in 2004 with the discovery of a two-dimensional electron gas (2DEG) at the interface with the band insulator LaAlO$_3$~\cite{Ohtomo2004}. Later, a similar 2DEG was observed at the bare surface of \STO{} where it is accessible to ARPES experiments~\cite{Meevasana2011a,Santander-Syro2011a}.
Catalysis and electronic structure studies are primarily interested in bulk terminated \STO. Preparing high quality bulk terminated surface remains a challenge though. \STO{} has no natural cleavage plane and surfaces prepared by sputtering and annealing show a plethora of surface reconstructions~\cite{Castell2002}.

Fig.~\ref{fig_STO}(b) shows a conventionally cleaved/broken \STO{} crystal imaged by optical profilometry. We find a smoothly curved surface 
with characteristically strong variations of the reflected light intensity. Such surfaces resembling broken glass are typical for conchoidal fracturing which occurs in brittle crystalline materials lacking a preferential cleavage plane.
In contrast, the micro-pillar in Fig.~\ref{fig_STO}(c) shows the optical properties of a high-quality cleave. With the exception of a few large steps and point-like defects, light is reflected in the same way from almost the entire surface area because of the prevalence of a single cleavage plane.
\textit{Ex-situ} atomic force microscopy (AFM) of cleaved \STO(001) micro-pillars shows long terraces bound by straight steps (Fig.~3(d,e)). Terrace widths vary between cleaves and over the surface area of a single cleave. In the representative areas imaged in Fig.~3(d,f), we find average terrace widths of 0.2~$\mu$m and 1~$\mu$m, respectively.
These values are large compared to the electronic mean free path. Terrace widths are thus not expected to limit the data quality obtainable in ARPES experiments. Step heights in these areas are around 10~nm and 2~nm, respectively corresponding to a miscut of $\approx 0.6^{\circ}$ in both cases. Such terraces directly confirm the effectiveness of the strain lense which guides the fracture along the desired cleavage plane and avoids instabilities leading to conchoidal fracturing.

Preliminary ARPES measurements on the \STO{} surface characterized in Fig.~\ref{fig_STO}(a-e) were performed at the high-resolution branch of beamline I05 at Diamond Light Source (DLS)~\cite{Hoesch2017}. 
While in general the micropillars are stable for transport to a beamline, one has to gain some experience in handling fragile samples and revisit the usual workflow accordingly to avoid accidental cleavage.
Inadvertedly, this occurred for the \STO(001) sample used for the ARPES measurements shown in Fig.~\ref{fig_STO}(f,g), which cleaved in the UHV system around 12 hours before the start of the beamtime, likely because of vibrations of a transfer arm. Hence, the ARPES data had to be taken on a surface that was already slightly contaminated before the start of the measurements.
Nevertheless, we observe multiple quantum well states of the well-known surface two-dimensional electron gas with a data quality comparable to the most successful of a large number of conventional cleaves~\cite{McKeown2015}. 
Similarly, first data on \STO(111) micro-pillars shown in Fig.~\ref{fig_STO}(h) is comparable in quality to the best data selected from multiple conventional cleaves~\cite{McKeown2014}.

In Fig.~\ref{fig_SRO} we show that FIB-machined micropillars are suitable for obtaining flat surfaces perpendicular to the natural cleavage plane of a layered perovskite. This offers new perspectives for ARPES as it gives ready access to $k_c$, the momentum component perpendicular to the layers. Sampling $k_c$ in traditional ARPES experiments on samples cleaved in-plane requires scanning of the photon energy. Such measurements require careful and time consuming energy calibrations and suffer from the intrinsically poor momentum resolution of ARPES along the surface normal~\cite{Hufner2003}. Their interpretation is further complicated by deviations from the commonly used free-electron final state approximation~\cite{Strocov2023}.
In contrast, on micro-cleaved surfaces perpendicular to the natural cleavage plane, the $k_c$ dispersion can readily be measured at a single photon energy by simple scanning of the emission angles using a sample goniometer or a deflector lens in the electron spectrometer.
Note that such measurements nevertheless profit from tunable synchrotron radiation, which allows selecting the desired plane in 3D $k$-space with a specific momentum component perpendicular to the cleaved surface.

Fig.~\ref{fig_SRO}(e) demonstrates the feasibility of this new approach on the example of a \SRO{} micro-pillar cleaved along the \textit{ac} plane.
\SRO{} has a low-temperature resistive anisotropy near $10^4$, far greater than the anisotropy of the relevant hopping integrals. This has been attributed to spin-orbit coupling induced anti-crossings that reduce the effective dimensionality of the Fermi surface~\cite{Haverkort2008}. However, direct measurements of the warping of the Fermi surface along $k_c$ proved difficult. Fig.~\ref{fig_SRO}(e) shows the first such direct measurement of the out-of-plane dispersion. Our data from the \textit{ac} surface of \SRO{} resolve two nearly straight Fermi surface contours corresponding to highly 2D Fermi surface sheets. 
For these measurements we tuned the photon energy to 83~eV, corresponding to $k_{\perp}\approx 6\pi/a$, so as to probe the $(k_a,0,k_c)$ plane in the 3. Brillouin zone. For this high-symmetry cut, the contribution of $k_{\perp}$ broadening vanishes to first order.
The good agreement of our experimental Fermi surface with a density functional theory calculation in the $(k_a,0,k_c)$ plane that includes spin-orbit coupling allows us to identify the experimental contours with the $\beta$ and $\gamma$ sheets, which dominate charge transport in \SRO.

\section{Outlook}
FIB shaped micropillars have the potential to advance the field in multiple directions. 
Controlled cleaving promises to improve the reproducibility and efficiency of ARPES and STM experiments, which greatly accelerates electronic structure studies. Microscopic control over the fracture process will further allow to select a cleavage plane within inhomogeneous materials, such as heterostructures or semi-conductor devices. Buried structures were so-far inaccessible to STM or ARPES, yet with FIB-prepared micropillars a desired plane of interest buried at depth can be selected as the cleavage plane. It will be interesting to explore how these techniques may provide additional insights into devices and complex structures.

FIB-micropillars will naturally also broaden the range of materials and surfaces accessible to these techniques, as we demonstrated on the $ac$-face of Sr$_2$RuO$_4$ (Fig.~4). 
Such micro-cleaves along planes containing the $c$-axis of layered crystals directly allow access to the out-of-plane dispersion of structurally layered materials, such as cuprate and iron-based high-T$_c$ superconductors. 
These measurements are largely immune to complications of the photoemission final states and will often be more precise than the traditional approach based on scanning of the photon energy because they are much less affected by the intrinsic final state broadening along the surface normal and do not require a precise calibration of the Fermi level for a large number of photon energies. The latter will be particularly valuable for measurements of the superconducting gap along $k_c$.
With reduced electronic dimensionality being a common thread in quantum materials, determining the strength of the electronic interlayer coupling is at the heart of understanding their physics. 
In the context of topological materials, surface state measurements on selected surfaces along multiple crystallographic directions are decisive for establishing their topological state~\cite{Noguchi2019}.

In some situations, a well-defined sample shape after the cleave is desirable. A topical case are experiments under directional strain, which crucially rely on a defined sample geometry after the cleave. FIB shaping permits the precise definition of cleavage planes, and hence allows an \textit{a priori} mechanical design of the cleaved sample.
These pioneering examples will undoubtedly become even more powerful in the future as the underpinning technology evolves. Firstly, this concerns advances in ARPES. With the advent of high-resolution ARPES beamlines with sub-10~$\mu$m beam diameter, it appears promising to further reduce the cleaved cross section, which will likely improve the surface quality. 
Such samples will no longer be compatible with the the simple manually actuated wobblesticks typically used to cleave samples at ARPES beamlines. Introducing suitable precision cleaving tools, preferably on the transferable sample plates, will be an important development of the method. It will also facilitate sample fabrication and handling by removing the need for a large top post. 

A second class of improvements are expected from current trends in FIB instruments, which moved away from being Ga-source dominated towards an ecosystem of multi-species sources that allows to select the right source and ion for a given task. The enhanced resolution of Ne-beams reduces the radius of curvature at the notch, enhancing the surface stress and localizing the cleavage plane more precisely. Utilizing the deep penetration depth of He ions may facilitate even more difficult cleaves by introducing bulk defects within the cleaving plane. Akin to the smart-cut utilized in semiconductors, generating deep defects can further weaken bonds at desired planes - a potential pathway towards nanometric positioning of the cleave in semiconductor devices. At the same time, high-current plasma-based sources are indispensable for this process, as large-scale structures have to be machined within acceptable time and cost.

It will be exciting to monitor the evolution of this highly versatile process in the future, when it adapts to match the requirements of more and more complex science questions.

\begin{acknowledgements}
We thank S. McKeown-Walker for discussions.
This work was supported by the Swiss National Science Foundation (SNSF) and by the European Research Council (ERC) under the European
Union’s Horizon 2020 research and innovation program (grant no. 715730, MiTopMat). We acknowledge Diamond Light Source for time on Beamline I05 under Proposal SI25083 and the Paul Scherrer Institut, Villigen, Switzerland for provision of synchrotron radiation beamtime at the SIS beamline of the SLS.
\end{acknowledgements}

\providecommand{\noopsort}[1]{}\providecommand{\singleletter}[1]{#1}%

\end{document}